\documentstyle[12pt,aaspp4,amssym,flushrt]{article}

\begin{document}

\title{Faint galaxy population in clusters:
X-ray emission, cD halos and projection effects.}

\author{Carlos A. Valotto\altaffilmark{1},
Hern\'an Muriel\altaffilmark{1},
Ben Moore\altaffilmark{2} and Diego G. Lambas\altaffilmark{1}}

\affil{val@oac.uncor.edu, hernan@oac.uncor.edu, ben.moore@durham.ac.uk,
dgl@oac.uncor.edu}

\altaffiltext{1}{IATE, Observatorio Astron\'omico,
Laprida 854, C\'ordoba and
CONICET, Argentina.}

\altaffiltext{2}{Institute for Theoretical Physics, University of Zürich,
Winterthurerstrasse 190, CH-8057 Z\"urich, Switzerland}

\begin{abstract}

We analyze samples of nearby
clusters taken from the Abell catalog and the X-ray Sample of Bright Clusters
(De Grandi et al 1999) including a wide range of X-ray luminosities.
Using the usually adopted
 background subtraction procedures, we
find that galaxies in clusters selected by means of their X-ray emission
show a flat luminosity
function (faint end slope $\alpha \simeq -1.1$) consistent with that derived
for galaxies in
the field and groups.  By contrast, the sample of Abell clusters that do not have an
X-ray counterpart shows a galaxy luminosity function with a steep faint end
($\alpha \simeq -1.6$).

We investigate the possibility that cD halos could be formed by the disruption of
galaxies in rich relaxed clusters that show an apparently
flat faint end galaxy
luminosity function (Lopez-Cruz et al 1997).
We find that clusters dominated by a central cD galaxy
(Bautz-Morgan classes I and II) show the same systematic trend: X-ray selected
clusters have flatter faint end slopes than those clusters with no detected X-ray emission.
Thus, it is likely the X-ray selection and not the cluster
domination by central galaxies what correlates with
background decontamination estimates of the
galaxy luminosity function. Moreover, no significant correlation between X-ray
luminosity and the galaxy LF faint end slope is found.
These results do not support a scenario where flat faint end slopes are a consequence of
cD formation via the disruption of faint galaxies. We argue that the clusters
without X-ray emission are strongly affected by projection effects which give rise to
spurious faint end slopes estimated using background subtraction procedures (Valotto et
al 2001).

\end{abstract}

\newpage

\section{Introduction}

Determining the galaxy luminosity function (hereafter LF) down to faint
magnitudes in rich clusters has been the subject of many studies in the last
years ({\it e.g.} Sandage, Binggeli \& Tammann 1985; Driver et al. 1994, De
Propris et al. 1995; Lobo et al.  1997; Lumsden et al. 1997; Lopez-Cruz et
al. 1997; Valotto et el. 1997; Wilson et al. 1997; Smith et al. 1997; Trentham
1998a; Trentham 1998b; Driver et al.  1998; Garilli et al. 1999).  Most of
these authors show that galaxy clusters are dominated by a large population of
high surface brightness dwarf galaxies, corresponding to a steep faint end
slope of the luminosity function, $\alpha$ in the range -1.4 to -2.0.

The origin and evolution of this faint galaxy population has important
consequences for our current understanding of galaxy formation and evolution
in dense environments.  For instance, Lopez-Cruz et al (1997) have suggested
that the galaxy LF is flat in dynamically evolved clusters characterized by
the presence of a dominant cD galaxy, high richness, symmetrical single-peaked
X-ray emission, and high gas mass. On the other hand, steep faint-end slopes
($-2.0 \leq \alpha \leq -1.4$) are detected in poorer clusters. It is worth noticing
the fact that the galaxy luminosity function of groups selected in redshift space is flat
at faint magnitudes (Muriel, Valotto \& Lambas, 1998) which shows the lack of a
universal trend of the parameter $\alpha$ with  system richness.
More recently, Martinez et al. (2002), have obtained reliable determinations
of galaxy LF in groups obtained from the 2dF galaxy redsfhit survey.
Their results are consistent with a nearly universal galaxy LF with little dependence on
environment which is consistent with these previous findings.
Lopez-Cruz et al. (1997)
suggest that cD galaxies are formed from the disruption of many faint galaxies
in the cluster cores, thus resulting in a globally flat faint end slope.
Indeed, dynamical processes operating in relaxed clusters are, in general,
destructive.  Ram pressure stripping (Gunn
\& Gott 1972, Abadi, Moore \& Bower 1999) and gravitational tides/galaxy
harassment ({\it e.g.} Moore et al. 1996) both tend to fade galaxies by
removing gas or stripping stars. These process are most effective for smaller,
less bound galaxies, and would cause a flattening of the faint end slope.
However, mergers and
galaxy interactions, when the cluster environment then was
more like a group environment could have contributed in the opposite direction
at early evolutionary stages.

Although large deep redshift surveys of
clusters are not presently available, Ferguson \&  Sandage (1991),
and Zabludoff \& Mulchaey (2000)
have analyzed cluster fields with redshift information and obtained faint end slopes
$\alpha \sim -1.25$ to $-1.3$, which are not significantly steeper than field
and group measurements in the 2dF galaxy redshift survey, Norberg et al. 2001,
Martinez et al. (2002) both consistent with $\alpha \sim -1.13$.
Most other estimates of galaxy luminosity functions in
clusters rely critically on background subtraction and are in general consistent with
significantly steeper faint end slopes ($\alpha \le -1.5$).
The accuracy of this procedure depends on the
statistical assumption that galaxy clusters correspond to density enhancements
unbiased with respect to the distribution of foreground or background
galaxies.  However, significant projection effects are found in Abell
clusters (e.g. Lucey 1983, Sutherland 1988, Frenk et al 1990) that can
systematically bias the observed correlation function and the mass function of
these systems.  In fact, van Haarlem et al (1998) claim that one third of Abell
clusters are not real physically bound systems but simply projections of
galaxies and groups along the line of sight.

Valotto, Moore and Lambas (2001) have analyzed several sources of systematic
effects present in observational determinations of the galaxy luminosity
function in clusters. They used mock catalogues derived from numerical
simulations of a hierarchical universe to identify clusters of galaxies in two
dimensions in a similar fashion to Abell 1958 and Abell et al. 1989.  Applying
standard background
subtraction procedures to these data gave rise to artificially steep faint end
slopes since many of the clusters do not have significant counterparts in
physical space. These projection effects result almost entirely from
the large scale structure behind the cluster, a result that was also concluded
by Adami et al (2000) from measuring $\sim 100$ redshifts for faint galaxies
thought to lie in the Coma cluster.
Color information (eg. b-r) is useful to improve the signal to noise
in the process of background decontamination by efficiently removing red
background galaxies at $z\geq0.5$. However, very unlikely
contaminating structures at significantly lower redshifts
would be eliminated by the use of colors. See for instance Adami et al. (2000)
where a significant number of background galaxies in the field of Coma Clusters
are at $0.02 < z < 0.3$.

The X-ray emission of the hot intracluster gas provides a confirmation of the
presence of a bound cluster of galaxies.  Thus, estimates of the
galaxy luminosity function in clusters by means of background decontamination techniques
restricted to an X-ray selected sample may provide a useful insight on the issues previously mentioned. In this paper we
explore the nature of the faint galaxy population in clusters
obtained by background decontamination techniques in X-ray and optically identified
clusters of galaxies.
By considering a subsample restricted to clusters dominated by
a central cD galaxy we can also explore the
disruption hypothesis suggested by Lopez Cruz et al (1997).

\section{Analysis}

\subsection{Cluster samples}
The sample of galaxy clusters used in our statistical analysis is taken from
Abell et al (1989) cluster catalog and from the X-ray flux-limited sample of
Bright X-ray Clusters (De Grandi et al 1999, hereafter BXS), an X-ray selected cluster sample
 based on the first analysis of the
ROSAT All-Sky Survey data (RASS1). This sample is count-rate
limited in the ROSAT hard band (0.5-2.0 keV) and its effective flux limits
varies between $\simeq$ 3 and 4 $\times10^{-12}$ ergs cm$^{-2}$ s$^{-1}$.

The region explored is limited to galactic latitude $b<-40\arcdeg$ and
declination $-70\arcdeg<\delta <-10\arcdeg$ and the area covered by the RASS1
Bright Sample. This survey area is restricted to regions with high exposure
time ($>$150s), excluding the sky areas of the Galactic plane and the
Magellanic clouds to avoid incompleteness in the cluster sample.  Due to the
lack of homogeneity of the sky coverage of the ROSAT All-Sky Survey, we have
checked the positions of those Abell clusters with no X-ray counterpart to
eliminate from the statistics objects in the poorly sampled regions (see
Figure 1).

\begin{figure}
\epsfxsize=0.5\textwidth
\plotone{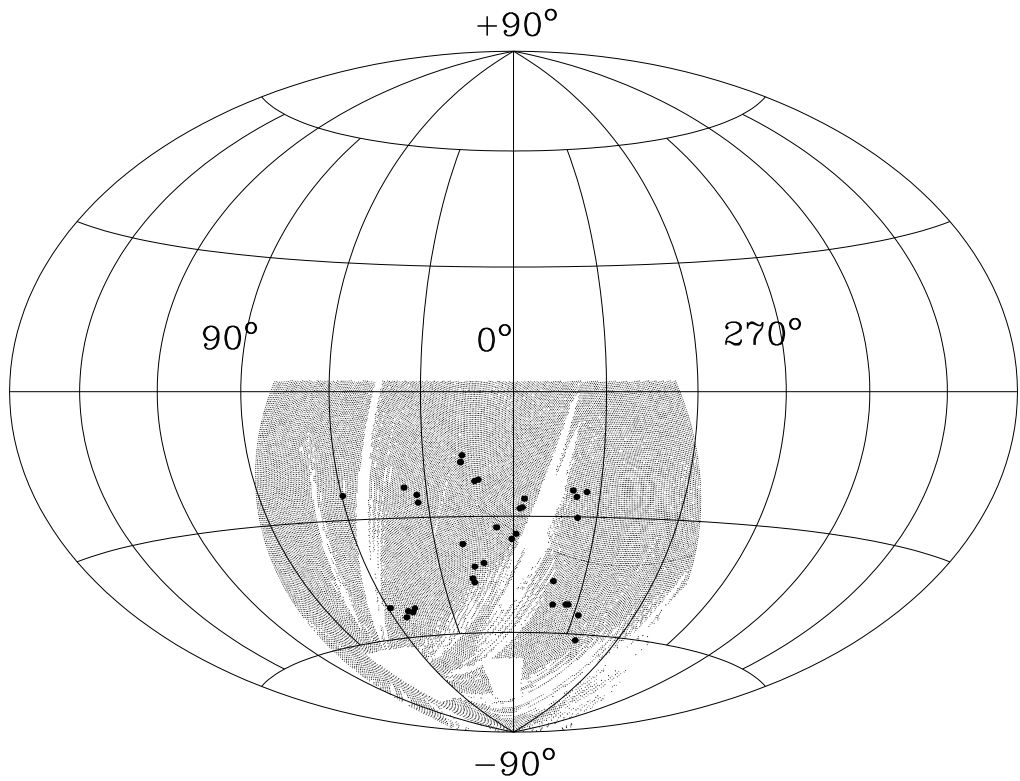}
\caption
{Projected distribution of Abell clusters and the area covered by
the ROSAT Bright Sample in equatorial coordinates (De Grandi et al., 1999)
\label{fig1}}
\end{figure}

In the area considered there are 34 Abell clusters and 15 X-ray clusters with
redshift $z<0.06$. Eight of these Abell clusters are identified with X-ray
emission in this sample.  Table 1 and table 2 list angular positions
(J2000.0), mean redshifts, richness number counts and Bautz-Morgan types for
the different samples analyzed here taken from Abell, Crowin \& Olowin 1989.
A low redshift cutoff is required in order
to reach faint absolutes magnitudes which unambiguously determine the $\alpha$
parameter.

\subsection {LF Determination}
The Edinburgh-Durham Southern Galaxy Catalog, hereafter COSMOS Survey
(Heydon-Dumbleton et al 1989), was used for the statistical assignment of
galaxies to the clusters. This survey provides angular positions and
photographic magnitudes in the $b_{j}$ band for over two million galaxies. We
have restricted to the region $\delta <-10 \arcdeg$ given the lower quality in
the photographic material in the northern hemisphere. We use a limiting
apparent magnitude, $b_j=20.0$ (Valotto et al 1997), for our analysis of the
COSMOS Survey which minimizes errors due to misclassification of stars,
galaxies plate variations, etc.
Incompleteness effects arise mainly due to star-galaxy
misclassification. In the COSMOS survey the latter
is expected to be lesser than 10\% at m=20.5,and completeness to be greater than 99.5 \%
at  $b_j$=19.5 (see Szapudi \& Gazta\~naga, 1998 for a comparison between COSMOS and APM
survey statistical properties).

The counts of galaxies for each cluster are binned in equal number intervals.
We subtract the corresponding mean background correction to each magnitude bin
to compute the contribution from each cluster to the LF.
We compute the number
of galaxies brighter than a limiting absolute magnitude $M_{\lim }$ within a
projected radial distance $r$ from the cluster centers.
The limiting absolute magnitude used is $M_{\lim }=-16.5$.
We have applied a
$K-$correction term of the form $K=3z$ (Efstathiou, Ellis \& Peterson 1988).
The projected radius $r$
is fixed at $ 1.0$ $h^{-1}$ $Mpc$. We assume the Hubble constant is $H_0=100~h$ km
$s^{-1}$ Mpc$^{-1}$, similar to that adopted in other studies. Since cluster redshifts are
very small (z<0.06) we simply use the local euclidean approximation.

We define a mean local background around each cluster in order to perform a
statistical background subtraction.  This mean local background is defined as
the number density of galaxies in the same range of apparent magnitudes in a
ring at projected radii $R_{1}<r<R_{2}$.
We have used $R_{1}=6$ Mpc h$^{-1}$ and $R_{2}=8$ Mpc h$^{-1}$.

According to Valotto et al. (1997)
the stability of the results does not depend crucially on the projected clusters
radii nor
on the adopted radius
for the background correction provided that the decontamination ring is well
beyond the average projected radius of the clusters and small enough in order to
take into account local variations of the projected galaxy density due to
patchy galactic obscuration, large scale gradients in the galaxy catalog, etc.

For all samples we compute error bars in the galaxy LF through bootstrap
resampling of the clusters to provide an estimate of the variations from
cluster to cluster.

In order to provided suitable fits for the galaxy luminosity functions, we have adopted
a Schechter function model
$\phi(L)dL=C\times (L/L^*)^{\alpha}e^{-L/L^*} d (L/L^*)$, where $C$ is a
constant (Schechter 1977).
We have applied a maximum likelihood estimator using the $\chi^2$-estimator procedure,
which minimize the difference

$$
\chi^2= \sum^N_{i=1} \left[ \phi_i - \phi(L_i;C, \alpha, L^*) \over \sigma_i\right],
$$
where $\phi_i$ is the relative frequency of galaxies corresponding to the $i$th
bin and $\sigma_i$ is its associated uncertainty.
All galaxy luminosity functions where arbitrarily normalized in order to make a proper
comparison of their shapes.

\section {Results}

In this section we discuss the results obtained from the analysis of our
cluster samples defined above.  In Figure 2 we show the galaxy LF for the
sample of bright X-ray clusters.  For comparison we show the LF for the sample
of Abell clusters in the same area of the sky and for the same
range of redshifts. The solid lines correspond
to Schechter function fits with parameters $\alpha=-0.9 \pm 0.1$, $M^*=-19.0
\pm 0.2$ and $\alpha=-1.50 \pm 0.1$, $M^*=-20.3 \pm 0.2$ for the X-ray and Abell
samples respectively. The X-ray defined cluster sample has a significantly flatter
faint end slope than the sample of Abell clusters.

\begin{figure}
\epsfxsize=0.5\textwidth
\plotone{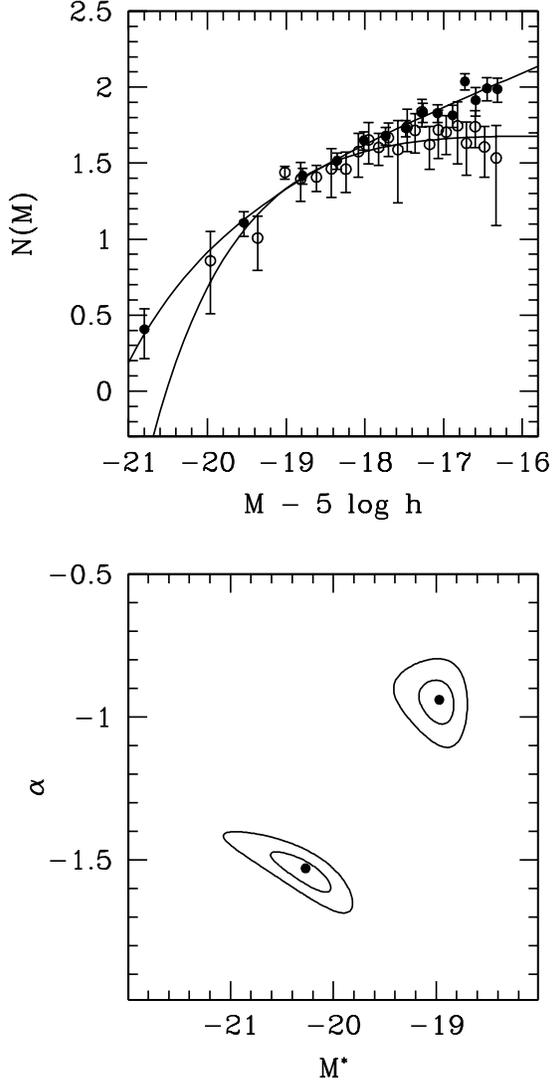}
\caption
{Galaxy luminosity function $N(M)$
for cluster galaxies of the Bright Cluster Sample (15 objects),
open circles,
and Abell clusters (35 objects), solid circles, corresponding to.
The solid line shows the best Schechter function fits for these samples:
$M^*=-19.0 \pm 0.1$, $\alpha= -0.9 \pm 0.1$
$M^*=-20.3 \pm 0.2$, $\alpha= -1.5 \pm 0.1$ respectively. The galaxy luminosity function is
computed within a projected radial distance from the cluster center $r = 1.0$ Mpc $h^{-1}$.
Error bars correspond to bootstrap resampling estimates.
\label{fig2}}
\end{figure}

We have also computed the galaxy LF for two subsamples of Abell clusters:
those confirmed by the X-ray intracluster emission, and those with no X-ray
detection. The results for these two samples are shown in Figure 3. Again we
find a clear difference in the LF of clusters with and without X-ray
emission.  Abell clusters with no detected X-ray emission show a very steep
galaxy LF faint end which contrasts with the flat behavior of the galaxy LF in
X-ray confirmed clusters. The corresponding Schechter fits are $\alpha=-1.0
\pm 0.1$, $M^*=-18.9 \pm 0.2$ and $\alpha=-1.6 \pm 0.1$, $M^*=-20.6 \pm 0.2$ for
the Abell clusters with and without X-ray detection respectively

\begin{figure}
\epsfxsize=0.5\textwidth
\plotone{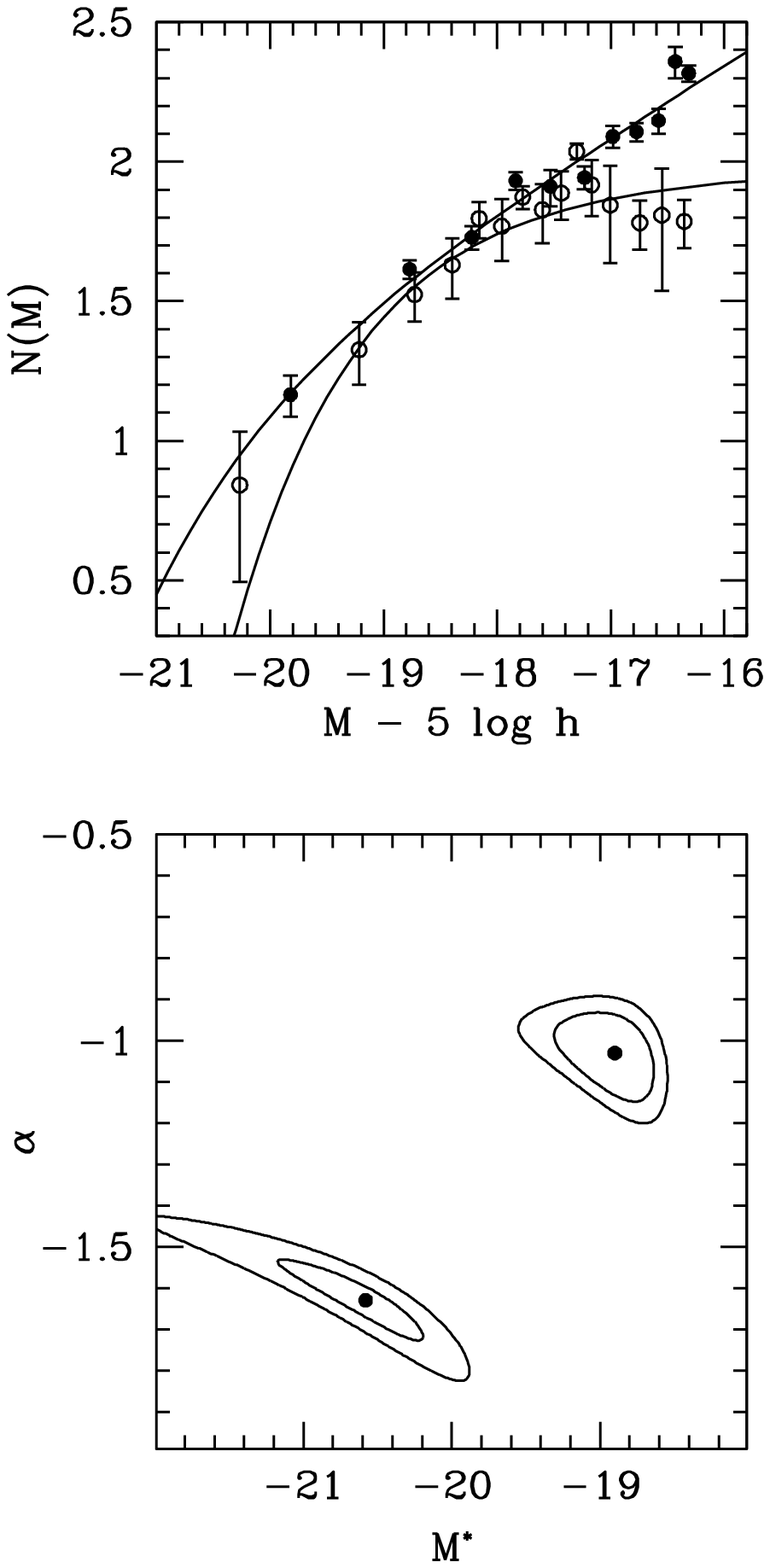}
\caption
{Galaxy luminosity function $N(M)$
for galaxies in clusters from the Bright Cluster Sample that are
also within the Abell catalog (9 objects),
open circles,
and Abell clusters with no X-ray counterpart (20 objects), solid circles.
The solid line shows the best Schechter function fits fort these samples:
$M^*=-18.9 \pm 0.2$, $\alpha= -1.0 \pm 0.1$
$M^*=-20.6 \pm 0.2$, $\alpha= -1.6 \pm 0.1$ respectively.
The projected radial distance $r$ and the
error bars are the same as in figure 2.
\label{fig3}}
\end{figure}

A main point in the analysis of Lopez-Cruz et al (1997) is the suggestion
that cD halos could be formed by accretion and disruption of
galaxies resulting in a flattening of the faint end slope of the LF.
Lopez-Cruz et al (1997) have suggested a scenario where the flat faint
end of the galaxy LF in relaxed clusters results from the disruption of
dwarf galaxies during the early stages of cluster evolution which also
may explain the halos of central cD galaxies  and a substantial fraction
of the intracluster medium.
We have tested this hypothesis
by computing the galaxy LF for a subsample of clusters with
Bautz-Morgan (BM) types I and I-II characterized by the presence of a
dominant cD galaxy (this restricts our analysis to Abell clusters for which
BM types are available).  The first subsample corresponds to
clusters selected in X-rays and the second subsample to clusters with no
detection of X-ray emission.  The results shown in Figure 4 suggests
that the BM type of the clusters does not determine the faint end galaxy
LF. Clusters with no detected X-ray emission have steep galaxy LF
, on the contrary, clusters with detected X-ray emission show flat galaxy
LF at the faint end.

\begin{figure}
\epsfxsize=0.5\textwidth
\plotone{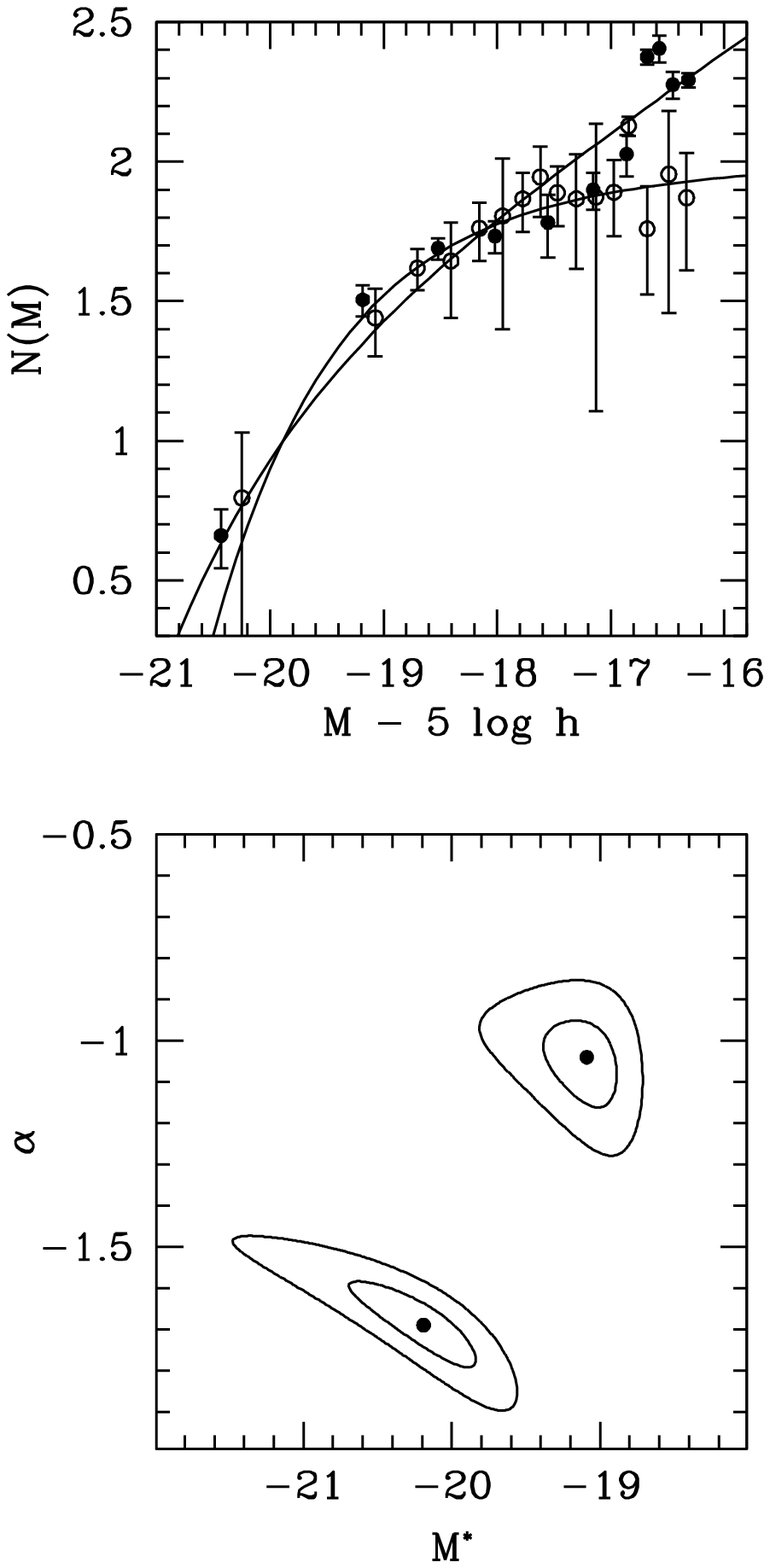}
\caption
{Galaxy luminosity function $N(M)$
for galaxies in clusters with Bautz-Morgan types I and I-II of the Bright Cluster Sample
with Abell identification
(6 objects), open circles,
and Abell clusters with no X-ray counterpart (10 objects), solid circles.
The solid line shows the best Schechter function fits fort these samples:
$M^*=-19.1 \pm 0.1$, $\alpha= -1.0 \pm 0.1$
$M^*=-20.2 \pm 0.2$, $\alpha= -1.7 \pm 0.1$ respectively.
The projected radial distance $r$ and the
error bars are the same as in figure 2.
\label{fig4}}
\end{figure}

The corresponding Schechter fits are $\alpha=-1.0 \pm 0.1$, $M^*=-19.1 \pm 0.1$
and $\alpha=1.7 \pm 0.2$, $M^*=-20.2 \pm 0.2$ for the X-ray and non x-ray
subsamples respectively.  This test indicate that the detection of a faint
population of faint galaxies is not correlated with dominant central galaxies
but with the X-ray confirmation.

In Lopez-Cruz et al. 1997 scenario, it would also be expected that the slope
of the faint end LF should correlate with the X-ray luminosity, i.e. the most
luminous X-ray clusters should have the flattest galaxy LF faint end slopes.
We explore this possibility by dividing our
X-ray cluster sample into equal numbers of clusters corresponding to
high and low X-ray luminosity ($L_x >1.0 \times 10^{44}$ ergs cm$^{-1}$
s$^{-1}$ and $L_x <1.0 \times 10^{44}$ ergs cm$^{-1}$ s$^{-1}$) which
correspond to mean luminosities of 2.44 and $0.83 \times 10^{44}$ ergs cm$^{-1}$ s$^{-1}$
respectively.

The resulting
LF's are shown in Figure 5 and
we find no differences between the two subsamples ($\alpha=-1.1 \pm
0.3$, $M^*=-18.6 \pm 0.2$ and $\alpha=-1.1 \pm 0.3$, $M^*=-18.8 \pm 0.2$ for low
and high X-ray luminosity sample respectively),
although the numbers of clusters in each sample is small.
This result indicates that
it is the detection of the intracluster medium through the X-ray emission
and not its
luminosity that correlates with the faint end slopes, giving support to the
hypothesis of strong projection contamination in optically identified cluster samples.

\begin{figure}
\epsfxsize=0.5\textwidth
\plotone{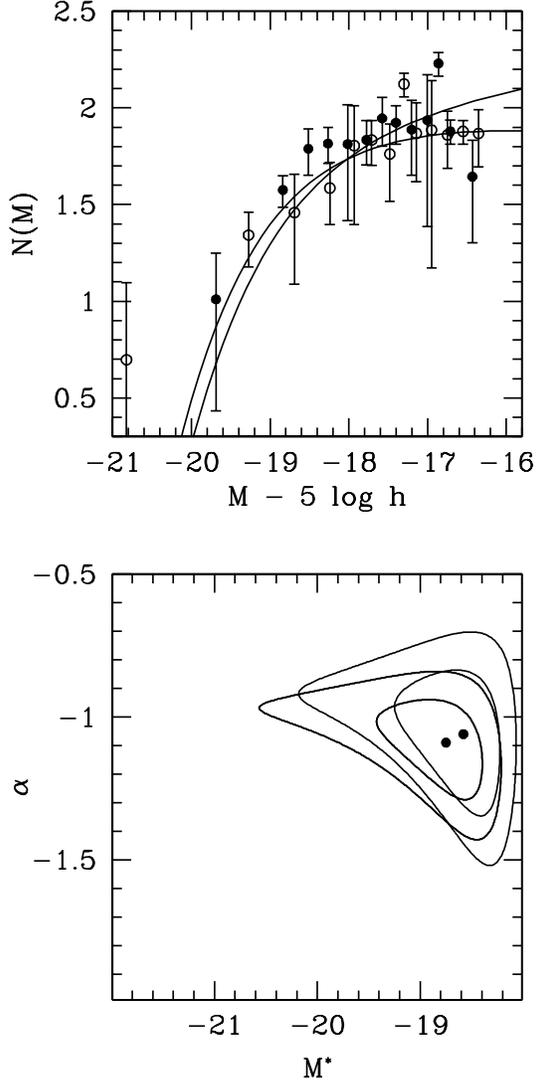}
\caption
{Galaxy luminosity function $N(M)$
for galaxies in clusters of the Bright Cluster Sample
with low X-ray luminosity ($L_x<1.0 \times 10^{44}$ ergs cm$^{-1}$ s$^{-1}$ ,
4 objects), open circles,
and high luminosity ($L_x>1.0 \times 10^{44}$ ergs cm$^{-1}$ s$^{-1}$,
5 objects), solid circles.
The solid line shows the best Schechter function fits fort these samples:
$M^*=-18.8 \pm 0.3$, $\alpha= -1.1 \pm 0.2$
$M^*=-19.6 \pm 0.3$, $\alpha= -1.1 \pm 0.2$ respectively.
The projected radial distance $r$ and the
error bars are the same as in figure 2.
\label{fig5}}
\end{figure}

We have analyzed samples of  clusters which no dominant central galaxies, i.e. BM types II,
II-III and III. We confirm here  that, again, X-ray detectability and not the presence or absence of dominant central galaxies correlates with the $\alpha$ parameter obtained from background decontamination.

In order to test far a possible dependence of our results on cluster radii, we have
analyzed all samples within two different limiting radii, $R = 0.5 h^{-1}$ Mpc and
$R = 1.0 h^{-1}$ Mpc. We find similar results for these two samples which indicates a lack of
strong radial dependence (see however de Propis et al. 1995).

Table 3 summarizes all of the
parameters for the Schechter function fits to the various cluster subsamples.
(We find no difference in our measurements of the faint end slope when we
constrain the value of $M^*$ to be the same value for each data set.)

We have also explored the ratio of dwarf to giant galaxies
(D/G, D: $-18.5<B_j<-16.5$, G: $-24<B_j<-18.5$)
in order to test the stability of our results
when Schechter functions give
poor fits to the actual LF.
We have computed the D/G ratios for all samples analyzed and we
show the results in
table 3.
As it can be seen by inspection to this table
D/G and $\alpha$ are strongly correlated,
all clusters with $\alpha > -1.2$ have D/G $< 4$
while samples with raising faint end slopes, $\alpha < -1.4$, have D/G $> 5$.
Thus, D/G ratios provide confidence on our faint end slope estimates.

\section {Discussion and Conclusions}

We have analyzed the galaxy LF
obtained by background decontamination techniques
in samples of nearby clusters taken from
Abell et al. 1989 and the Bright X-ray Cluster Catalogues (De Grandi et al.
1999). We find that X-ray selected clusters show  an apparently flat
luminosity function consistent with that derived for the field and groups($\alpha \simeq -1$).  By contrast,
clusters of galaxies identified from the projected galaxy distribution
that do not have an X-ray counterpart show a galaxy luminosity function with a
very significant steeper faint end ($\alpha \simeq -1.5$).

We find that for the subsample of clusters with dominant central galaxies
(Bautz-Morgan type I and I-II) the shape of the faint end galaxy LF depends on
the detection of the X-ray emission of the intracluster gas.  In fact, we derive a steep
galaxy LF for the subsample of clusters with central dominant
galaxies with no detected X-ray intracluster emission.  This fact argues
against the hypothesis that the disruption of faint galaxies would provide the
material out of which cD halos form causing a flattening of the faint end
slope.
A note of caution should be set here since we have
used Bautz-Morgan classes as a
suitable division between cD and non-cD clusters.
Although some BM type I and I-II
clusters may not contain bonafide cD galaxies,
these clusters are strongly dominated by a central galaxy,
so it is on the global, statistical sense that our analysis provide a
test of Lopez-Cruz et al. hypothesis.

Our results could be influenced by the possibility that
many X-ray undetected
clusters could be bound systems less
dynamically evolved and therefore with a large fraction of
emission line galaxies which have a steeper
alpha than non-emission line galaxies (Madgwick et al. 2002).
Nevertheless,
Martinez et al. 2002 have shown that even low mass groups ($M_{virial}\sim 10^{13}
-10^{14} M_{\odot}$) in the 2dF galaxy redshift survey  are dominated by absorption line
type galaxies suggesting that this is not a very serious possibility.

More likely, our results provide support to the presence of biases on the
cluster galaxy LF
derived by background decontamination procedures
due to projection effects, as suggested by Valotto,
Moore and Lambas (2001). Clusters identified from the projected galaxy distribution
are biased by  many spurious clumps with no physically bound system along the line
of sight. The resulting luminosity functions from background decontamination procedures
show  steep faint end
slopes and can be erroneously interpreted as the clusters being dominated
by a population of dwarf galaxies.
Furthermore, the fact that no significant correlation between X-ray
luminosity and the galaxy LF faint end slope is found
argues against processes associated to the gaseous environment causing the
differences in the galaxy LF faint end slope.

\section{Acknowledgments}
We thank the anonymous Referee for very helpful suggestions
which greatly improved the previous version of this paper.
This research was supported by grants from
Agencia C\'ordoba Ciencia, Secretar\'{\i}a
de Ciencia y T\'ecnica
de la Universidad Nacional de C\'ordoba, Fundaci\'on Antorchas,
and Agencia Nacional de Promoci\'on Cient\'{\i}fica,
Argentina.
CV is supported
by the National Science Foundation
through grant \#GF-1003-00 from the Association of Universities for Research
in Astronomy, Inc., under NSF cooperative agreement AST-9613625.

\clearpage

\newpage

\begin{deluxetable}{cccccc}
\tablecaption{Sample of Abell Clusters with no X-ray identification}
\tablewidth{0pt}
\tablehead{
\colhead{Name} &
\colhead{$\alpha$} &
\colhead{$\delta$} &
\colhead{$z$} &
\colhead{$Abell Richness$} &
\colhead{$BM$}
}
\startdata
A2731 & 00 10.2 & -56 59 & 0.0312    & 39 & III  \\
A2806 & 00 40.2 & -56 09 & 0.0275    & 37 & I-II  \\
A2824 & 00 48.6 & -21 20 & 0.0486    & 46 & III  \\
A0114 & 00 53.7 & -21 40 & 0.0580    & 43 & -  \\
A2870 & 01 07.7 & -46 54 & 0.0250    & 33 &I \\
A2877 & 01 09.8 & -45 54 & 0.0231    & 30 &I \\
A2881 & 01 11.2 & -17 04 & 0.0445    & 36 &II \\
A2882 & 01 11.4 & -17 04 & 0.0455    & 32 &II\\
A2896 & 01 18.3 & -37 06 & 0.0317    & 44 &I \\
A2933 & 01 40.7 & -54 33 & 0.0208    & 77 &III \\
A2992 & 02 14.9 & -26 40 & 0.0584    & 30 &I \\
A2995 & 02 15.2 & -24 50 & 0.0370    & 69 &I-II \\
A0419 & 03 08.5 & -11 31 & 0.0410    & 32 &- \\
A3093 & 03 10.9 & -47 23 & 0.0635    & 93 &I \\
A3125 & 03 27.4 & -53 30 & 0.0590    & 46 &III \\
A3144 & 03 37.1 & -55 01 & 0.0430    & 54 &I-II \\
A3193 & 03 58.2 & -52 20 & 0.0345    & 41 &I \\
A3816 & 21 50.3 & -55 17 & 0.0389    &  39&I-II \\
A3782 & 21 34.5 & -62 01 & 0.0557    &  40&II \\
A3851 & 22 16.7 & -52 35 & 0.0520    &  33&I-II \\
A3869 & 22 21.4 & -52 34 & 0.0396    &  49&II \\
A3893 & 22 38.0 & -23 54 & 0.0330    & 39 &I \\
A3925 & 22 51.8 & -46 35 & 0.0510    & 38 &II \\
A4012 & 23 31.8 & -33 49 & 0.0510    & 35 &II-III \\
A4013 & 23 31.9 & -35 16 & 0.0500    & 51 &III \\
A2660 & 23 45.3 & -25 58 & 0.0520    & 45 &- \\
\enddata
\end{deluxetable}

\begin{deluxetable}{ccccccc}
\tablecaption{Bright X-ray Cluster subsample}
\tablewidth{0pt}
\tablehead{
\colhead{Name} &
\colhead{$\alpha$} &
\colhead{$\delta$} &
\colhead{$z$} &
\colhead{$Abell Richness$} &
\colhead{$BM$} &
\colhead{$L_x [10^{44} erg~ s^{-1}]$}
}
\startdata
A4038 & 23 47.7 & -28 08 & 0.0290    & 117 &III&1.01 \\
A4059 & 23 56.7 & -34 40 & 0.0470    & 66 &I& 1.79\\
A3911 & 22 46.1 & -52 43 & 0.0381    & 58 &II-III& 2.43\\
A3880 & 22 27.8 & -30 34 & 0.0581    &  31&II& 0.90\\
A3158 & 03 42.9 & -53 38 & 0.0591    & 85 &I-II&3.36 \\
A0133 & 01 02.6 & -21 47 & 0.0604    & 47 & - & 2.16 \\
A0151 & 01 08.9 & -15 25 & 0.0536    & 72 &II& 0.77\\
A2717 & 00 03.3 & -35 57 & 0.0498    & 52 &I-II  &0.64 \\
S0041 & 00 25.6 & -33 03 & 0.0566 &  &  & 0.68\\
0118.5-1408 & 01 21.0 & -13 51 & 0.0511 &  &  &1.03 \\
S0301 & 02 49.6  & -31 11 & 0.0223 &  &  &0.10 \\
1ES0412-382 &  04 14.0 & -38 06 & 0.0502  &  &  &0.84 \\
ESO235-G050 &  21 04.8 & -51 49 & 0.0491 &  &  &0.88 \\
ESO146-G028 &  22 29.0 & -60 54 & 0.0412 &  &  &0.24 \\
S1101 & 23 14.0 & -42 44 & 0.0580 &  &  &1.93 \\
\enddata
\end{deluxetable}
\newpage

\begin{deluxetable}{ccccc}
\tablecaption{Results}
\tablewidth{0pt}
\tablehead{
\colhead{Sample Description} &
\colhead{Number of cluster} &
\colhead{$\alpha$} &
\colhead{$M^*$} &
\colhead{D/G}
}
\startdata
Abell              & 34 & $-1.5\pm0.1$ & $-20.3\pm0.2$ & $5.45\pm0.16$ \\
BXS                & 15 & $-0.9\pm0.1$ & $-19.0\pm0.1$ & $3.20\pm0.16$ \\
Abell non-X        & 26 & $-1.6\pm0.1$ & $-20.6\pm0.2$ & $6.46\pm0.22$ \\
Abell X            & 8  & $-1.0\pm0.1$ & $-18.9\pm0.2$ & $3.94\pm0.25$ \\
Abell I, I-II non-X& 12 & $-1.7\pm0.1$ & $-20.2\pm0.2$ & $5.74\pm0.31$ \\
Abell I, I-II X    & 3  & $-1.0\pm0.1$ & $-19.1\pm0.1$ & $3.87\pm0.26$ \\
BXS,  low X        & 4  & $-1.1\pm0.2$ & $-18.8\pm0.3$ & $3.34\pm0.30$ \\
BXS,  high X       & 4  & $-1.1\pm0.2$ & $-18.6\pm0.3$ & $3.84\pm0.42$ \\
\enddata
\end{deluxetable}

\end{document}